\documentclass[english,aps,prl,superscriptaddress,preprint]{revtex4}

\usepackage{amsmath}
\usepackage{graphicx}
\usepackage{amssymb}

\makeatletter
\usepackage{ae,aecompl}
\usepackage{subfigure}

\usepackage{babel}
\makeatother
\begin{document}

\title{Unexpected Density Fluctuations in Jammed Disordered Sphere Packings}

\author{Aleksandar Donev}

\affiliation{\emph{Program in Applied and Computational Mathematics}, \emph{Princeton
University}, Princeton NJ 08544}

\affiliation{\emph{PRISM, Princeton University}, Princeton NJ 08544}

\author{Frank H. Stillinger}

\affiliation{\emph{Department of Chemistry}, \emph{Princeton University}, Princeton
NJ 08544}

\author{Salvatore Torquato}

\email{torquato@electron.princeton.edu}

\affiliation{\emph{Program in Applied and Computational Mathematics}, \emph{Princeton
University}, Princeton NJ 08544}

\affiliation{\emph{PRISM, Princeton University}, Princeton NJ 08544}

\affiliation{\emph{Department of Chemistry}, \emph{Princeton University}, Princeton
NJ 08544}

\begin{abstract}
We computationally study jammed disordered hard-sphere packings as
large as a million particles. We show that the packings are saturated
and hyperuniform, i.e., that local density fluctuations grow only
as a logarithmically-augmented surface area rather than the volume
of the window. The structure factor shows an unusual non-analytic
linear dependence near the origin, $S(k)\sim\left|k\right|$. 
In addition to exponentially damped oscillations seen in liquids, this
implies a weak power-law tail in the total correlation function, $h(r)\sim-r^{-4}$,
and a long-ranged direct correlation function. 
\end{abstract}
\maketitle

\newcommand{\Cross}[1]{\left|\mathbf{#1}\right|_{\times}}

\newcommand{\CrossL}[1]{\left|\mathbf{#1}\right|_{\times}^{L}}

\newcommand{\CrossR}[1]{\left|\mathbf{#1}\right|_{\times}^{R}}

\newcommand{\CrossS}[1]{\left|\mathbf{#1}\right|_{\boxtimes}}

\newcommand{\V}[1]{\mathbf{#1}}

\newcommand{\M}[1]{\mathbf{#1}}

\newcommand{\D}[1]{\Delta#1}

\newcommand{\sV}[1]{\boldsymbol{#1}}

\newcommand{\sM}[1]{\boldsymbol{#1}}

\newcommand{\grad}{\boldsymbol{\nabla}}

The characterization of local density fluctuations in many-particle
systems is a problem of great fundamental interest in the study of
condensed matter, including atomic, molecular and granular materials.
In particular, long-wavelength density fluctuations are important to such diverse fields as
thermodynamics, granular flow, and even cosmology (see Ref. \cite{Torquato_Fluctuations} and references therein).
Previous work by some of us \cite{Torquato_Fluctuations} was concerned
with the quantitative characterization of density fluctuations in
point patterns, and in particular, those in which infinite wavelength
fluctuations are completely suppressed, i.e., the structure factor
$S(\V{k})$ vanishes at the origin. In these so-called \emph{hyperuniform}
(or \emph{superhomogeneous} \cite{HZ_Spectrum_rspace}) systems, the
variance in the number of points inside a large window grows slower
than the volume of the window, typically like the window surface area.
Most known examples of hyperuniform systems are either ordered lattices
or quasi-crystals \cite{Torquato_Fluctuations,HZ_Spectrum_rspace}.
An important open problem is the identification of statistically homogeneous
and isotropic atomic systems (e.g., glasses) that are hyperuniform.

For equilibrium liquids and crystals, $S(k=0)$ is proportional to
the isothermal compressibility and is thus positive. 
\emph{Strictly} \emph{jammed} sphere packings \cite{Torquato_jammed} are rigorously incompressible (and non-shearable)
\cite{Cones_strain}, but they are also nonequilibrium systems.
In Ref. \cite{Torquato_Fluctuations} it was conjectured that all
\emph{saturated} \cite{foot_1} strictly jammed packings are hyperuniform.
Of particular importance to understanding both glasses and granular materials
are disordered jammed packings, and in particular the maximally random jammed (MRJ) state \cite{Torquato_MRJ},
which is the most disordered among all strictly jammed packings \cite{Torquato_jammed}.
The MRJ state for hard-particle packings is related to the view of jamming
as a rigidity transition and/or dynamic arrest in both granular \cite{JammedMatter_Review}
and glassy materials \cite{ForceChains_OHern}. Previous studies have
identified several different diverging length scales at the rigidity
jamming transition for systems of soft spheres (see Ref. \cite{VibrationalSpectrum_Jamming} and references therein),
indicating a kind of second-order phase transition at the jamming point.
Hyperuniformity involves an {}``inverted critical phenomenon'' in which
the range of the direct correlation function $c(r)$ diverges \cite{Torquato_Fluctuations}. 
It is therefore of great interest to test whether disordered jammed sphere packings are hyperuniform.
In this Letter, we demonstrate that MRJ packings are indeed hyperuniform and saturated.
Moreover, we observe an unusual non-analytic structure factor $S(k)\sim\left|k\right|$,
or equivalently, a quasi-long ranged negative tail of the total pair
correlation function $h(r)\sim-r^{-4}$, just as found in the early Universe \cite{HZ_Spectrum_rspace}.

We prepare jammed packings of hard spheres under periodic boundary
conditions using a modified Lubachevsky-Stillinger (LS) packing algorithm
\cite{LS_algorithm}, as detailed in Ref. \cite{Jamming_g2}. The
generated disordered sphere packings typically have volume fractions
in the range $\phi=0.64-0.65$, and to a good approximation the packings
should be representative of the MRJ state. 
For this study, we have generated a dozen packings
of $N=10^{5}$ and $N=10^{6}$ particles jammed up to a reduced pressure
of $10^{12}$ using an expansion rate of $10^{-3}$ \cite{Jamming_g2}
with $\phi\approx0.644$. Generating such unprecedented one-million-particle
packings was neccesary in order to study large-scale density fluctuations
without relying on dubious extrapolations.

\begin{figure}
\begin{center}\includegraphics[%
  width=0.80\columnwidth,
  keepaspectratio]{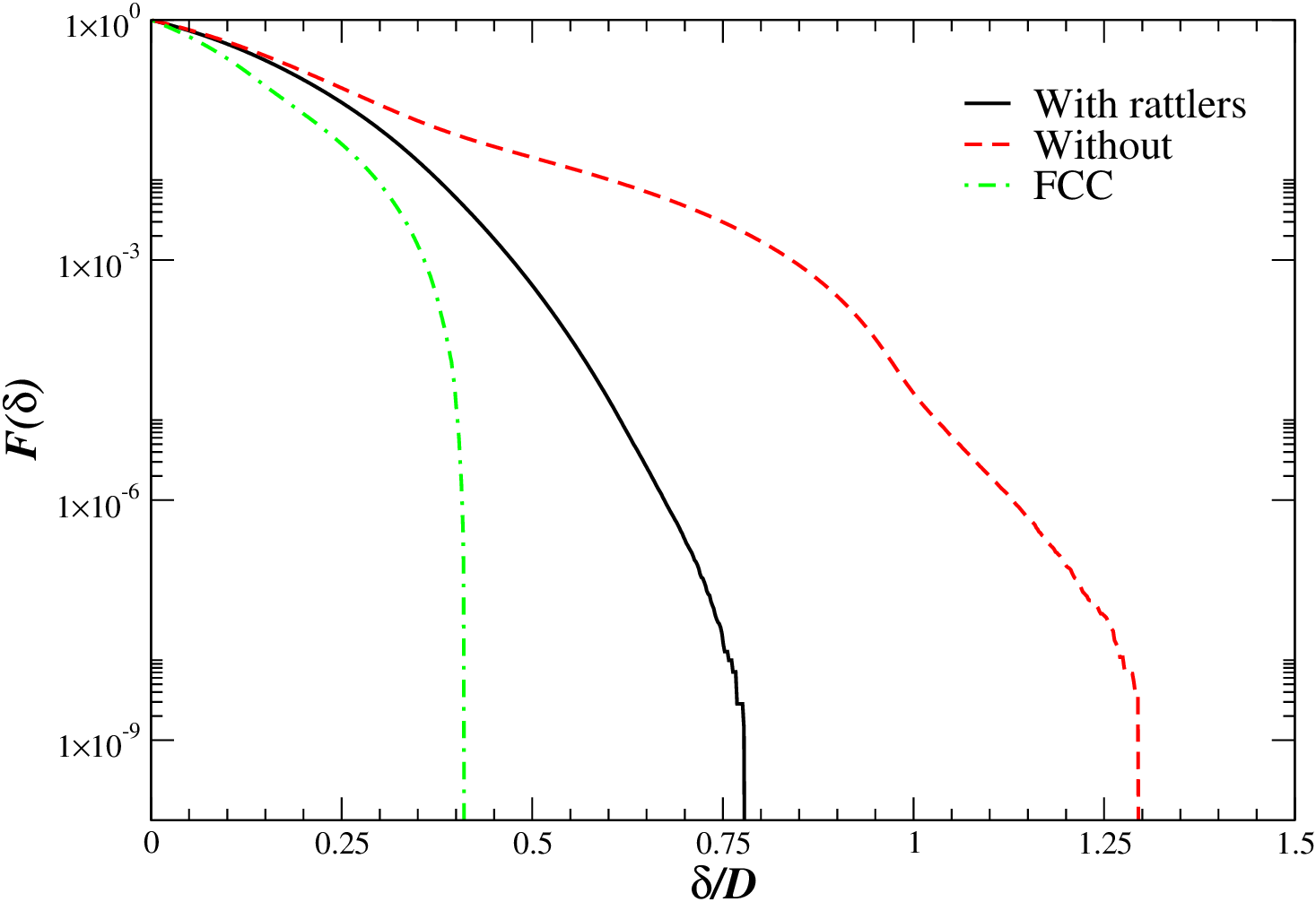}\end{center}

\caption{\label{Pore.HS.N=3D1E5}(Color online) The cumulative pore-size distribution $F(\delta)$
for a (single) packing with $N=10^{5}$ particles, with and without the rattlers.
The method of trial spheres with $2\cdot10^{9}$ trials was used \cite{Random_Materials}.
A very similar $F(\delta)$ with cutoff around $\delta_{\textrm{max}}\approx0.8D$
is observed when $N=10^{6}$, when rattlers are present. The cutoff
is however not as sharply defined as it is for the FCC crystal, shown
for comparison, since the exact size of the largest available void (cavity) fluctuates between realizations.}
\end{figure}

The packings generated by the LS and other algorithms have a significant
fraction ($\sim2.5$\%) of \emph{rattling} particles which are not
truly jammed but can rattle inside a small cage formed by their jammed
neighbors \cite{Jamming_g2}. These rattlers make a negligible contribution
to the mechanical properties of the system, including the pressure,
and can be removed sufficiently close to the jamming point. However,
they are important when considering density fluctuations. Removing
the rattlers will produce small but observable long-wavelength density
fluctuations. Assuming that the rattlers are more or less randomly
distributed among all particles, a hyperuniform packing from which
the rattlers are then removed would have $S(0)\approx0.025>0$. 
Similarly, the hyperuniformity could be destroyed by randomly filling large-enough
voids with additional rattlers. It is therefore important to verify
that the jammed packings are saturated, i.e., that there are no voids
large enough to insert additional rattlers. Figure \ref{Pore.HS.N=3D1E5}
shows the complementary cumulative pore-size distribution \cite{Random_Materials}
$F(\delta)$, which gives the probability that a sphere of diameter
$\delta$ could be inserted into the void space, with and without
the rattlers. Clearly there is no room to insert any additional rattlers;
the largest observed voids are around $\delta_{\textrm{max}}\approx0.8D$.
The algorithm used to produce the packings appears to fill all void
cages with particles.

It is very difficult to study long-wavelength
density fluctuations accurately in 3D computer simulations. When periodic
boundary conditions apply with a periodic box of length $L$, particle
correlations can only be studied up to a distance $L/2$, and there
are large finite-size corrections for distances comparable to $L$.
Additionally, as we show later, strong statistical fluctuations appear
due to finite system size, making it necessary to use even larger
systems to measure pair correlations at large distances. In reciprocal
space, $S(k)$ can only be measured for $k\geq2\pi/L$, with large
discretization errors for the smallest wavevectors. To overcome these
finite-size effects, it was necessary to generate a packing of one
million particles.

Consider a large isotropic three-dimensional packing of $N$ hard
spheres of diameter $D$, with average number density $\rho=N/V$
and average volume fraction $\phi=\pi\rho D^{3}/6$. We employ the
usual pair correlation function $g_{2}(x=r/D)$ or the total correlation
function $h(x)=g_{2}(x)-1$ in real space, or the equivalent Fourier
representation given by the structure factor \[
S(K=kD)=1+24\phi\int_{0}^{\infty}\frac{\sin\left(Kx\right)}{Kx}x^{2}h(x)dx.\]
Of particular interest are the moments of $h(x)$, $\left\langle x^{n}\right\rangle =\int_{0}^{\infty}x^{n}h(x)dx$.
Computer-generated packings are always finite, and thus binning techniques
must be used to obtain probability densities like $h$. Accordingly,
we prefer to use the more readily measurable \emph{excess coordination}
\[
\D{Z}\left(x\right)=1+24\phi\int_{0}^{x}w^{2}h(w)dw.\]
This is the average excess number of points inside a spherical window
of radius $xD$ centered at a particle, compared to the ideal-gas
expectation $8\phi x^{3}$. Any integral containing $h(x)$ can easily
be represented in terms of $\D{Z}\left(x\right)$ using integration
by parts. For the structure factor we get $S\left(K\right)=\lim_{R\rightarrow\infty}S(K,R)$,
where\begin{equation}
S(K,R)=\D{Z}\left(R\right)\frac{\sin\left(KR\right)}{KR}-\int_{0}^{R}\D{Z}\left(x\right)\frac{d}{dx}\frac{\sin\left(Kx\right)}{Kx}dx.\label{S_k_from_dZ}\end{equation}
This has quadratic behavior near $k=0$ when expanded in a Taylor
series,\begin{equation}
S\left(K\right)\approx S(0)+\frac{K^{2}}{3}\int_{0}^{\infty}x\left[\D{Z}(x)-S(0)\right]dx,\label{S_k_quadratic}\end{equation}
where $S(0)=\D{Z}\left(x\rightarrow\infty\right)$ vanishes for a
hyperuniform system. For large $x$, an \emph{explicit} finite-size
correction of order $1/N$ needs to be applied to the infinite-system
excess coordination, $\D{Z}(x)\approx S(0)\left[1-8\phi x^{3}/N\right]$
\cite{FiniteSize_Combined}, as it is clear that the excess coordination
must vanish for windows as large as the whole system.

\begin{figure}
\begin{center}\includegraphics[%
  width=0.95\columnwidth,
  keepaspectratio]{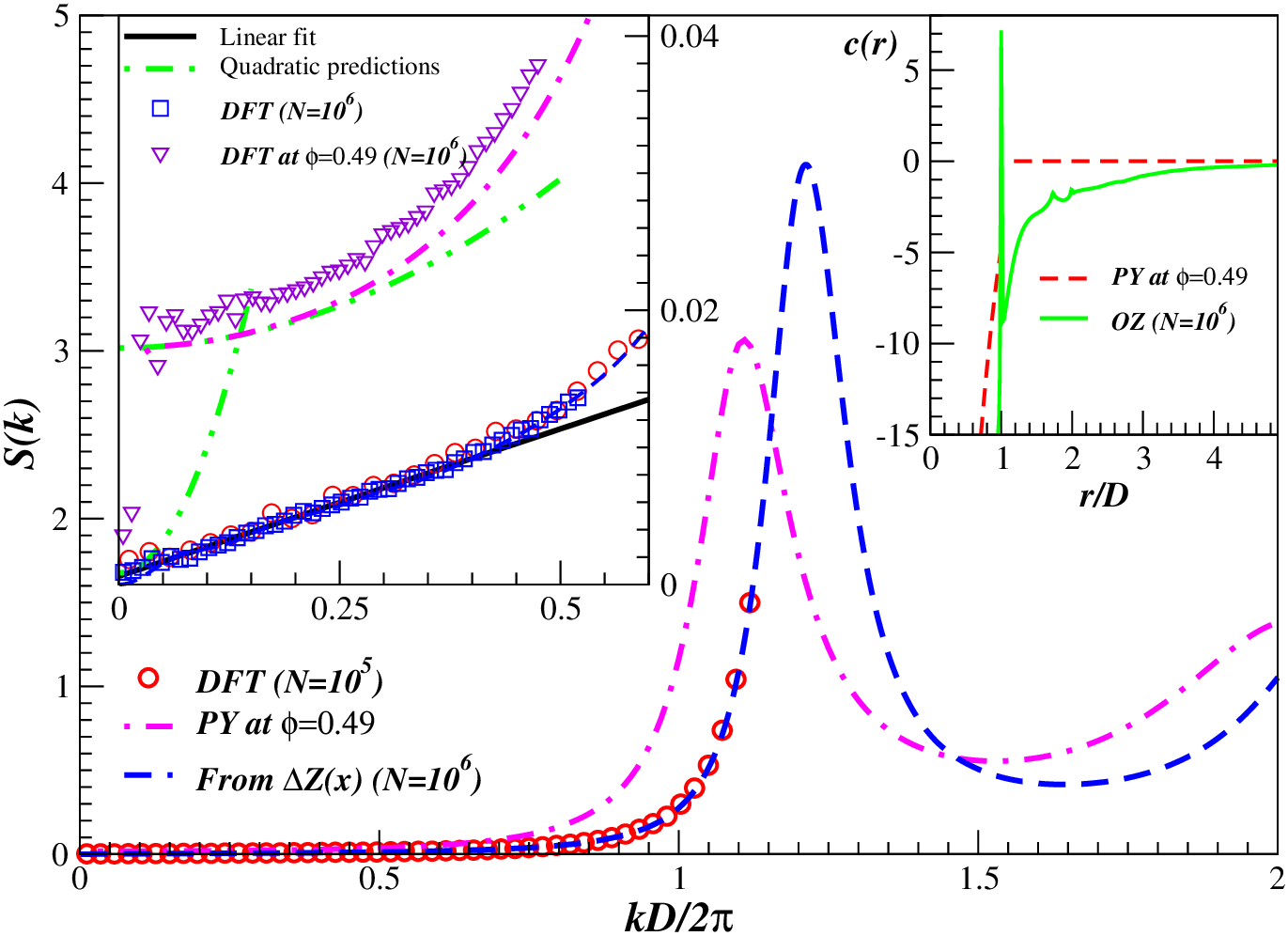}\end{center}

\caption{\label{S_k.HS.N=3D1E6}(Color online) Structure factor for a $10^{6}$-particle
packing ($\phi=0.642$) and for a hard-sphere liquid near the freezing point ($\phi=0.49$),
as obtained via two alternative numerical methods and also from the
Percus-Yevick (PY) theory for the liquid \cite{PY_original}. DFT
results are also shown over a larger range of $K$ for a $10^{5}$-particle
packing ($\phi=0.643$). The left inset focuses on the range close to the origin,
showing that while a parabola matches the liquid data reasonably well
{[} $S(K)\approx0.02+4\cdot10^{-3}K^{2}$ according to PY theory, which is known to underestimate $S(0)$ {]},
it does not appear appropriate for the jammed packing for large-to-intermediate
wavelengths [as obtained from Eq. (\ref{S_k_quadratic})]. 
The very linear behavior of the DFT data for the jammed
packing in the range up to $K/2\pi<0.4$ is remarkable. The right
inset shows $c(r)$ convolved (smudged) with a narrow Gaussian {[}due
to numerical truncation of $S(k)${]}. The peak at $r=D$ is thus
in fact (almost) a $\delta$-function.}
\end{figure}

Figure \ref{S_k.HS.N=3D1E6} shows $S(k)$ for the simulated packings
as obtained via a direct Fourier transform (DFT) of the particle positions,
$S(\V{k})=N^{-1}\left|\sum_{i=1}^{N}\exp\left(i\V{k}\cdot\V{r}_{i}\right)\right|^{2}$,
where $\V{k}$ is a reciprocal lattice vector for the periodic unit
cell \cite{foot_2}. To obtain an approximation to the radially symmetric infinite-system
$S(k)$, we average over the reciprocal lattice vectors inside a spherical
shell of thickness $2\pi/L$. Using Eq. (\ref{S_k_from_dZ}) together
with a numerical (truncated) $\D{Z}(x)$ quickly gives $S(k)$ over
a wide range of wavelengths. However, the behavior near the origin
is not reliable since it depends on the analytic extension for the
tail of $\D{Z}(x)$. The results of the DFT calculations are shown
in Fig. \ref{S_k.HS.N=3D1E6}, and they closely match the one obtained
from $\D{Z}(x)$ for wavelengths smaller than about 20 diameters.

Figure \ref{S_k.HS.N=3D1E6} demonstrates that the saturated packing
is indeed hyperuniform \cite{foot_3} to within $S(0)<10^{-3}$, as conjectured in Ref. \cite{Torquato_Fluctuations}.
The behavior of $S(k)$ near the origin is very surprising. The observed
$S(k)$ follows closely a \emph{non-analytic} \emph{linear} relationship
\cite{foot_4} well-fitted by $S(K)\approx6.1\cdot10^{-4}+3.4\cdot10^{-3}K$ over
the whole range $K/2\pi<0.4$. By contrast, analytic quadratic behavior
is observed for a liquid sample at $\phi=0.49$, as shown in the figure.
Theoretical finite-size corrections to the small-$k$ behavior of
$S(k)$ have only been considered for relatively low-density liquid
systems with relatively small $N$ \cite{FiniteSize_Combined}, and
do not appear useful for our purposes. Although estimating the corrections
to the DFT data analytically is certainly desirable, such corrections
appear to be rather small at least for the well-understood liquid
at $\phi=0.49$. Comparison among the different $N=10^{6}$ samples
shows that statistical fluctuations in $S(k)$ near the origin are
very small.

Equation (\ref{S_k_quadratic}) shows that if $h$ is truly short-ranged,
the structure factor must be analytic (i.e., an even power of $k$,
usually quadratic), near the origin. Our numerical observations point
strongly to a linear $S(K)$ for small $K$. It is interesting to
note that such non-analytic behavior is assumed in the so-called Harrison-Zeldovich
power spectrum of the density fluctuations in the early Universe \cite{HZ_Spectrum_rspace},
and has been verified experimentally with high accuracy. If this observation
$S(K)\sim\left|K\right|$ survives simulations of even larger jammed
hard sphere systems, using a variety of packing algorithms, it would
imply a negative algebraic power-law tail $h(x)\sim-x^{-4}$ uncharacteristic
of liquid states. Such a quasi-long range correlated decrease in the
density around a particle is typically only seen in systems with long-range
interactions. A long-ranged tail must appear in the \emph{direct}
correlation function $c(r)$ for a strictly hyperuniform system due
to the divergence of $\tilde{c}(0)$, in a kind of {}``inverted critical
phenomenon'' \cite{Torquato_Fluctuations}.
Such a tail is uncharacteristic
of liquids where the range of $c(r)$ is substantially limited to
the range of the interaction potential. The direct correlation function
can numerically be obtained from its Fourier transform via the Ornstein-Zernike
(OZ) equation, $\tilde{c}(k)=\left(\pi/6\phi\right)\left[S(k)-1\right]/S(k),$
and we have shown it in the inset in Fig. \ref{S_k.HS.N=3D1E6}, along
with the corresponding Percus-Yevick (PY) anzatz \cite{PY_original} for $c(r)$ at $\phi=0.49$ which
makes the approximation that $c(r)$ vanishes outside the core.
Two unusual features relative
to the liquid are observed for our jammed packing. First, there is
a positive $\delta$-function at contact corresponding to the $\bar{Z}=6$
average touching neighbours around each jammed particle \cite{Jamming_g2}.
Second, there is a relatively long tail outside the core, the exact
form of which depends on the behavior of $S(k)$ around the origin \cite{foot_5}.

\begin{figure}
\begin{center}\includegraphics[%
  width=0.85\columnwidth,
  keepaspectratio]{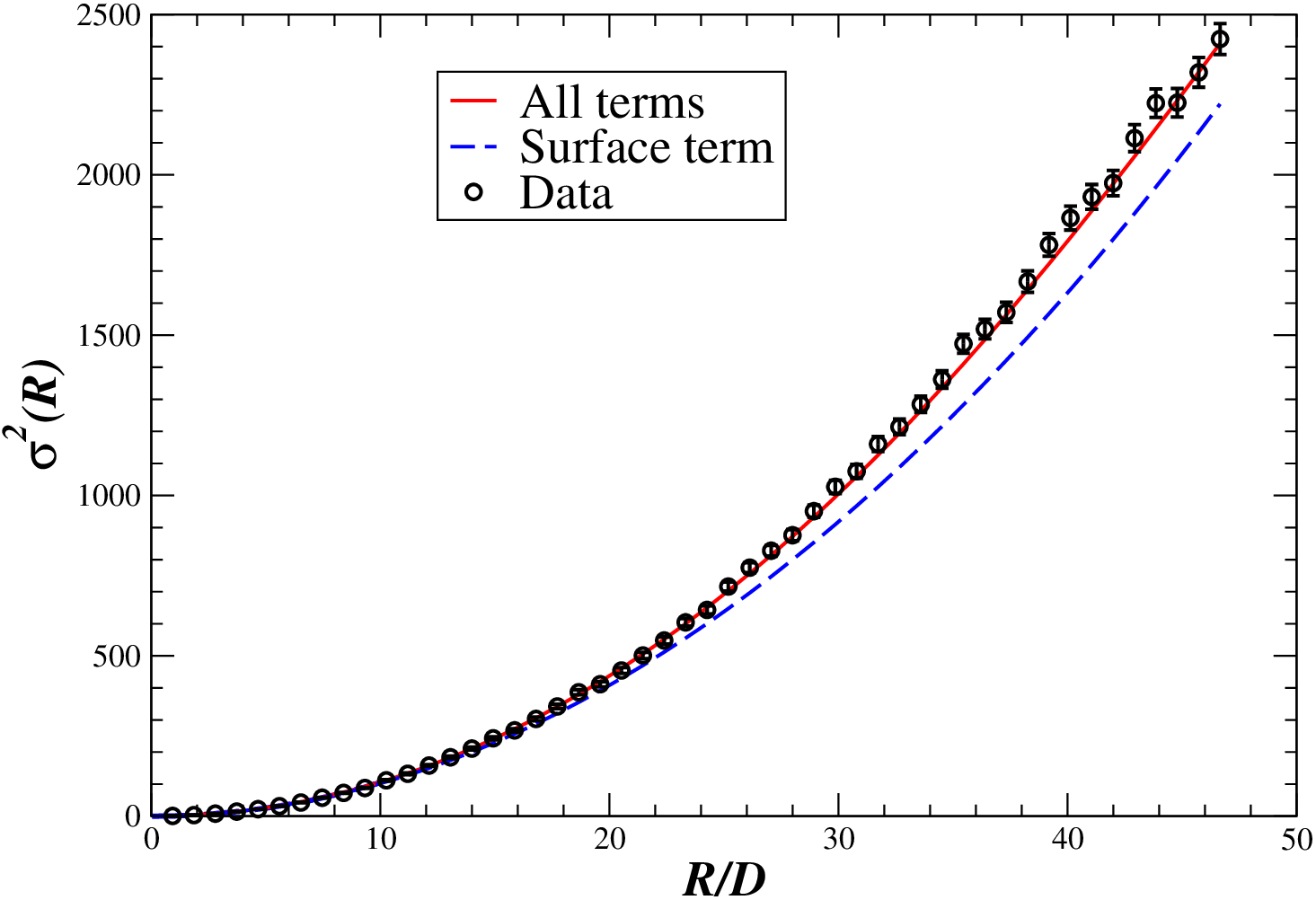}\end{center}

\caption{\label{dN2.HS.N=3D1E6}(Color online) The variance $\sigma^{2}$
as a function of the window radius for a $10^{6}$-particle packing.
The uncertainty in the variance, as shown with error bars, is estimated
to be of the order of $\sigma^{2}/\sqrt{M}$, where $M=10^{4}$ is
the number of windows used for a given window. Also shown is the theoretically
predicted dependence of the form $AX^{3}+CX^{2}\ln X+B_{0}X^{2}$,
along with just the surface term $B_{0}X^{2}$, which dominates the
density fluctuations.}
\end{figure}

The numerical coefficient in the power-law tail in $h(x)$ is very
small, $\Delta Z(x)\approx4.4\cdot10^{-3}x^{-1}$, and cannot be directly
observed, as we will show shortly. It is however possible to observe
its effect on large-scale density fluctuations. For monodisperse hard
sphere systems it suffices to focus only on the positions of the sphere
centers and consider density fluctuations in \emph{point patterns}.
Following Ref. \cite{Torquato_Fluctuations}, consider moving a spherical
window of radius $R=XD$ through a point pattern and recording the
number of points inside the window $N(X)$. The number variance is
exactly \cite{Torquato_Fluctuations},\begin{eqnarray*}
\sigma^{2}(X) & = & \left\langle N^{2}(X)\right\rangle -\left\langle N(X)\right\rangle ^{2}\\
 & = & \frac{3\phi}{2}\left[(2X)^{2}\D{Z}_{0}(2X)-\D{Z}_{2}(2X)\right]\end{eqnarray*}
where $\D{Z}_{n}(x)=\int_{0}^{x}w^{n}\D{Z}(w)dw$ denotes a running
moment of $\D{Z}$. Asymptotically, for large windows, in an infinite
system with analytic $S(k)$, $\sigma^{2}(X)\approx AX^{3}+BX^{2}$,
where $A=8\phi\left(1+24\phi\left\langle x^{2}\right\rangle \right)=8\phi S(0)$
is the \emph{volume} fluctuation coefficient, and $B=-144\phi^{2}\left\langle x^{3}\right\rangle =6\phi\D{Z}_{0}\left(x\rightarrow\infty\right)$
is the \emph{surface} fluctuation coefficient. When a non-integrable
power-law tail exists in $\D{Z}(x)$, asymptotically the {}``surface''
fluctuation coefficient contains an additional logarithmic term, $B(X)=B_{0}+C\ln X$.
Such a logarithmic correction does not appear for any of the examples
studied in Ref. \cite{Torquato_Fluctuations}. Explicit finite-size
effects for non-hyperuniform systems yield a correction $A(X)=8\phi S(0)\left[1-8\phi X^{3}/N\right]$
\cite{Fluctuations_Disks_Combined}. Figure \ref{dN2.HS.N=3D1E6}
shows numerical results for the number variance as a function of window
size, along with the predicted asymptotic dependence, including both
the logarithmic and $N^{-1}$ corrections \cite{foot_6}. 
Both corrections need to be included in order to observe this close
a match between the data and theory. The constants $S(0)$ and $C$
were obtained from the linear fit to $S(k)$, while $B_{0}\approx1.02$
was obtained by numerically integrating $\D{Z}(x)$, as explained
shortly \cite{foot_7}. 

\begin{figure}
\begin{center}\includegraphics[%
  width=0.95\columnwidth,
  keepaspectratio]{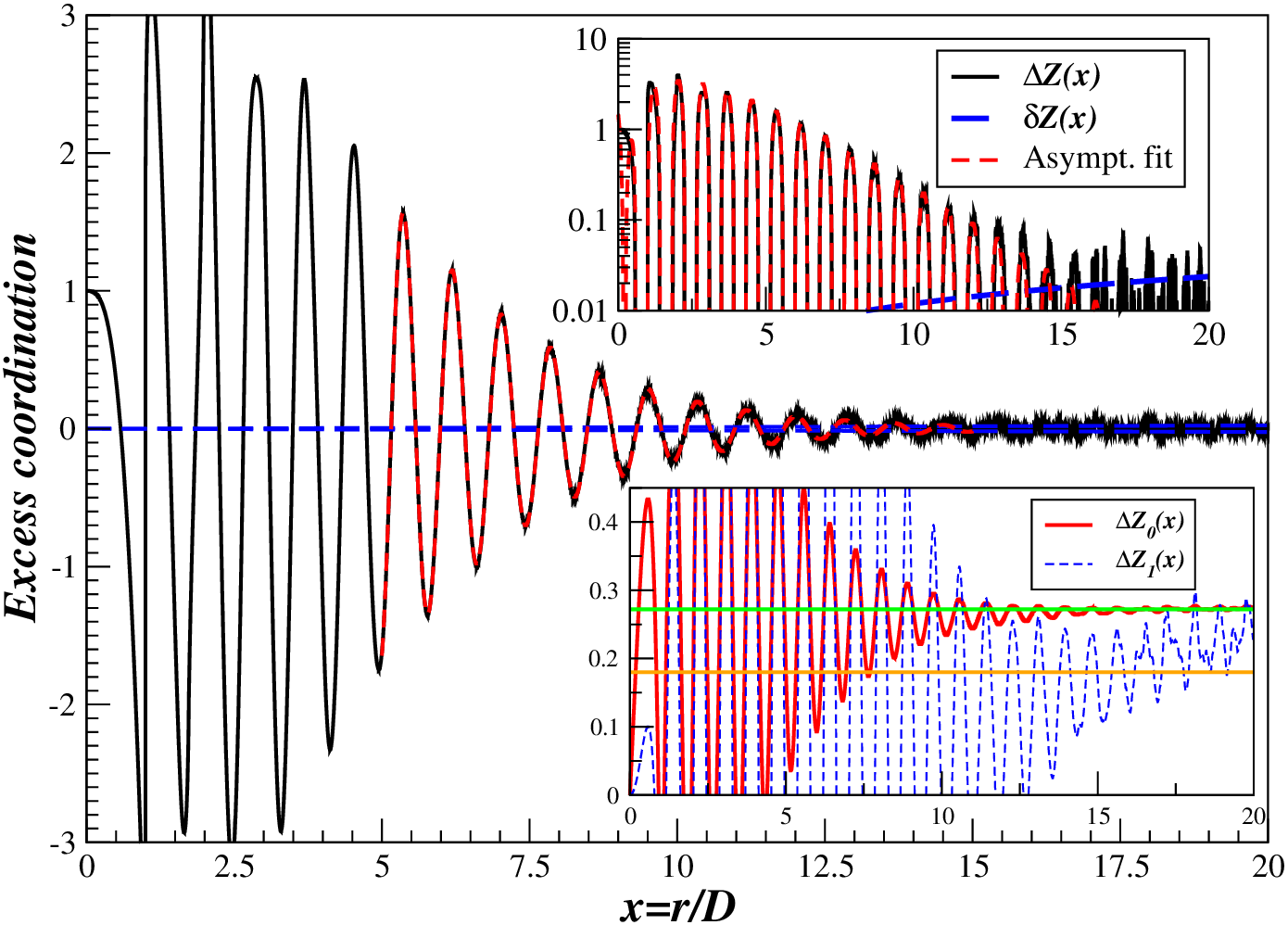}\end{center}

\caption{\label{dZ.HS.N=3D1E6}(Color online) The excess coordination for
a $10^{6}$-particle packing, along with the best fit of the form
(\ref{h_exp_oscillatory}) for the tail, and the estimated uncertainty.
Statistical fluctuations overcome the actual correlations after $x\approx15$.
Averaging over nine samples only shrinks the magnitude of the fluctuations
by three, without revealing additional information. The inset on the
top uses a logarithmic scale, and the inset on the bottom shows the
zeroth and first running moments along with their asymptotic values
as estimated from the tail fit. Note that for the range of $x$ shown
explicit finite-size corrections are small (less than 5\%).}
\end{figure}

We now turn our attention to real space to observe directly the large-distance
behavior of $h$ or equivalently $\D{Z}$. For equilibrium liquids
with short-ranged potentials it is expected that the asymptotic behavior
of $h(x)$ is exponentially damped oscillatory \cite{Decay_g2_PY,Torquato_trial_g2},
of the form\begin{equation}
h(x)\sim\frac{C}{x}\exp\left(-x/\xi\right)\cos\left[K_0(x-x_{0})\right].\label{h_exp_oscillatory}\end{equation}
 However, it is not clear whether the decay is still exponential for
glass-like nonequilibrum jammed systems. Previous studies have looked
at much smaller systems, where explicit finite-size effects dominate,
and also focused on the liquid phase \cite{FiniteSize_Combined}.
Figure \ref{dZ.HS.N=3D1E6} shows the numerical $\D{Z}(x)$ along
with a relatively good exponentially damped oscillatory fit \cite{RCP_JT_Footnote}
$\D{Z}(x)\approx5.47x\exp(-x/1.83)\cos(7.56x-2.86)$ over the range
$5<x<15$. It would be desirable to look at larger $x$ and, in particular,
directly observe the long-range inverse power tail predicted from
the linear behavior of $S(k)$. The use of cubic periodic boundary
conditions implies that pair distances up to $x_{\max}=\sqrt[3]{\pi N/24\phi}\approx50$
can be studied. However, it is important to point out that it is not
possible to measure the pair correlations for $x>15$ due to statistical
variations among finite systems, estimated to lead to an uncertainty
of the order $\delta Z(x)\approx\sigma(x)/\sqrt{N}$. In fact, within
the range of validity of the observed $\D{Z}(x)$ the damped oscillatory
fit is perfectly appropriate. We smoothly combined the actual numerical
data for $x<10$ with the fitted decaying tail for $x>10$, and numerical
integration of this smoothed $\D{Z}(x)$ gives $B_{0}\approx1.02\pm0.02$,
as used in producing Fig. \ref{dN2.HS.N=3D1E6}. This smoothed $\D{Z}(x)$
was used to obtain $S(k)$ via Eq. (\ref{S_k_from_dZ}) when producing
Fig. \ref{S_k.HS.N=3D1E6}.

We have given computational results for a million-particle jammed
disordered hard sphere packing demonstrating that it is nearly hyperuniform
and saturated. However, there are many open fascinating questions.
Can a geometrical significance be attached to the period of oscillations
$K_0$ in the jamming limit \cite{foot_8}, or to the cutoff of $F(\delta)$?
We believe that the strict jamming and saturation conditions demand
hyperuniformity of our packings. 
We conjecture that the observed non-analytic behavior of $S(k)\sim k$ 
is a direct consequence of the condition of maximal disorder on the jammed packing.
The exponent $p$ appears to increase with increasing order: It approaches infinity for ordered lattices, 
is two for perturbed lattices, and is one for MRJ. In Ref. \cite{Torquato_Fluctuations} we examined
the possibility of using the surface term coefficient $B$ as an order metric (increasing $B$
indicated greater disorder). We did not anticipate the appearance of a further logarithmic term 
for the disordered packings. In this sense, the MRJ packings are markedly more disordered: they have
macroscopic density fluctuations which are much larger than crystalline packings. Quantitative understanding
of this aspect of disorder and its relation to density fluctuations remains a fascinating open problem.

\begin{acknowledgments}
The authors were supported in part by the National Science Foundation under Grant No. DMS-0312067.
\end{acknowledgments}


\begin{thebibliography}{18}
\bibitem{foot_1}
A saturated packing is one in which no additional particles can be added.
\bibitem{foot_2}
This calculation potentially involves many reciprocal lattice points
and can only be done for a limited range of wavevectors near the origin
due to memory and CPU constraints, and cannot easily and accurately
be made faster using Fast Fourier Transforms (FFT) \cite{NFFT_Tutorial}.
\bibitem{foot_3}
We believe that this small residual non-hyperuniformity is an artifact
of the packing algorithm. For example, by reducing the particle expansion
rate one observes an increase in the hyperuniformity (and also in
density \cite{Jamming_g2}).
\bibitem{foot_4}
Since $S(k)$ is an even function, its derivative must vanish at the
origin for it to be analytic.
\bibitem{foot_5}
We used the linear fit to $S(k)$ when producing the figure, implying
$c(r)\sim r^{-2}$.
\bibitem{foot_7}
The surface coefficient $B_{0}$ cannot be determined from just the
linear part of $S(k)$ near the origin.
\bibitem{foot_8}
For example, the height of a tetrahedron of side $D$ is around $0.816D$, close to
the period of oscillations in $h$.
\end{thebibliography}
\end{document}